\documentclass[a4paper,12pt]{svjour2}
\usepackage{a4,pstricks,pst-node,amsmath,amssymb}
\usepackage[english]{babel}
\usepackage{amsmath}
\usepackage[latin1]{inputenc}
\usepackage[T1]{fontenc}
\usepackage{ae,aecompl}
\usepackage[hang]{caption}

\usepackage{graphics}
\usepackage{graphicx}

\usepackage{color}
\definecolor{darkblue}{rgb}{0,0,0.6}
\definecolor{darkred}{rgb}{0.6,0,0}

\usepackage[colorlinks=true,urlcolor=darkblue,citecolor=darkblue,linkcolor=darkred,hyperfootnotes=false,hyperindex,breaklinks]{hyperref}
\PassOptionsToPackage{linktocpage}{hyperref}
\usepackage{url}

\newcommand{\C}{\mathcal C}
\newcommand{\cc}{{\text{c}}}
\newcommand{\mk}{{\mathbb K}} \newcommand{\sym}{{\text{sym}}}
\renewcommand{\phi}{\varphi}

\title{Finite size scaling of the dynamical free-energy in a kinetically constrained model}
\date{\today}
\author{Thierry Bodineau \and Vivien Lecomte \and Cristina Toninelli}

\institute{
Thierry Bodineau 
\at \'Ecole Normale Sup\'erieure, DMA, 45 rue d'Ulm 75230 Paris cedex 05, France,
\and 
Vivien Lecomte
\and
Cristina Toninelli
\at Laboratoire Probabilit\'es et Mod\`eles Al\'eatoires, UMR CNRS 7599, 
 Universit\'es Paris VI et Paris VII, site Chevaleret, 175 rue du Chevaleret, 75013 Paris, France 
}

\begin{document}

\maketitle

\begin{abstract}
  We determine the finite size corrections to the large deviation
  function of the activity in a kinetically constrained model (the
  Fredrickson-Andersen model in one dimension), in the regime of
  dynamical phase coexistence. Numerical results agree with an
  effective model where the boundary between active and inactive
  regions is described by a Brownian interface.
\end{abstract}

\setcounter{tocdepth}{2}
\tableofcontents

\newpage

\section{Introduction}
\label{sec:intro}

Glassy phenomena have proven difficult to understand: they present a
variety of features --~slow dynamics, ageing, dynamical heterogeneity,
frustration~-- which make their study arduous from a theoretical point
of view (see~\cite{berthier_theoretical_2011} for a recent review).
Kinetically Constrained Models (KCMs) are a simple class of lattice
gases whose dynamics shares features similar to those of glassy
phenomena, with the advantage that no disorder is present in the model
--~which makes them easier to study
(see~\cite{ritort_glassy_2003,garrahan_kinetically_2010} for reviews
on KCMs).

There is a variety of KCMs (see section~\ref{subsec:ldfFA} for a
concrete example) which all share a common feature: their static
properties are trivial and their complexity (like the phase transition
phenomena) is hidden in their dynamical behaviour.  This raises the
problem of finding relevant physical parameters in order to describe
and classify the properties of these models.  As the glassy systems
are characterised by a mixture of frozen and mobile areas, the
``activity" of the system (namely the number of local updates during a
time interval) has been proposed as a relevant dynamical parameter and
a dynamical approach has been recently devised in order to define a
suitable notion of dynamical free energy
\cite{merolle_spacetime_2005,lecomte_thermodynamic_2007,garrahan_dynamical_2007,garrahan_first-order_2009,bodineau_toninelli_2011}.
In this dynamical framework, the role of the free energy is played by
the large deviation function of the activity.

\subsection{The large deviation function and its singularities}

For lattice gases (or more generically for Markov process
with discrete configuration space) with continuous-time dynamics, the
simplest definition of the activity is the number of configuration
changes presented by an history of duration $t$
~\cite{merolle_spacetime_2005,maes_wynants_2008}
(see~\cite{hedges_dynamic_2009,bodineau_lefevere_2008,pitard_dynamic_2011} for alternative definitions in
systems with continuous degree of freedom).
For  each history of duration $t$ of the system, the activity will be denoted by the observable $K_t$.
KCMs and other glassy systems present ``dynamical
heterogeneities''~\cite{chandler_dynamics_2010}, \emph{i.e.} regions which remain frozen
during a long time. 
This feature can be captured by the probability distribution
function of the activity as some histories
present slow or inactive intervals with higher probability than in
non-glassy systems. 
In the large-time
limit, the probability of observing an atypical value $K_t=kt$ of the activity scales as
\begin{equation*}
  \text{Prob}[K_t=kt] \underset{t\to\infty}\sim e^{t\pi(k)}
\end{equation*}
In the infinite size limit, the function $\pi(k)$ may no longer be analytic.
This can be interpreted as a signature of the  dynamical
heterogeneities~\cite{merolle_spacetime_2005,jack_space-time_2006}.
We postpone a more  quantitative
discussion of the singularities to section~\ref{subsec:pi_L_of_k_finite-size}.

From a practical point of view, it proves in fact easier to make a
Laplace transform and consider
instead~\cite{lecomte_thermodynamic_2007}
\begin{equation*}
  \big\langle e^{-sK_t}\big\rangle \underset{t\to\infty}\sim e^{t\psi(s)}
\end{equation*}
where the average is taken over histories of duration $t$.
The parameter $s$ plays a role similar to the inverse temperature in
the canonical ensemble of equilibrium statistical mechanics: fixing
$s$ boils down to fixing the average value of the activity, in the
same way as fixing the temperature determines the average energy.  The
functions $\psi(s)$ and $\pi(k)$ are related by a Legendre transform:
$\psi(s)=\sup_{k} \{ \pi(k)-sk \}$.  

It has been shown for several KCMs that $\psi(s)$ presents a
singularity at $s\downarrow 0$ in the infinite size limit, which
corresponds to a dynamical phase coexistence between active histories
(at $s\leq 0$) and inactive histories (at $s>0$)
\cite{garrahan_dynamical_2007,garrahan_first-order_2009},
in the same way as singularities of the free energy correspond 
to transition phases in equilibrium statistical mechanics.
Similar singularities have been observed in other glassy
systems (see~\cite{hedges_dynamic_2009,pitard_dynamic_2011}
for binary Lennard-Jones mixtures),
but the question of finding a generic relation between glassy properties
which hold at $s =0$ and the singularity of $\psi(s)$  is still open.
Indeed, for any finite size system  the cumulants of the activity: $\frac 1t
\langle K_t^n\rangle_c= (-1)^n \partial_s^{n}\psi(s)|_{s=0}$ can be obtained 
from the function $\psi(s)$, but this correspondence does not hold in the 
infinite size limit and one may wonder if the singularity of $\psi$ at $s=0$ 
has an impact for the physics of finite size dynamics.
Thus it is a natural question to understand how this singularity is build up when
the system size diverges and 
in this article we are interested in the finite-size scaling of the
large deviation function $\psi(s)$, especially around the transition.

\subsection{Finite size scaling of the large deviation function}

It has been shown that finite-size effects capture non-trivial
physical features of the typical configurations of the system giving rise
to the atypical deviation,  such as the stationarity or the stability of the density
profile in one-dimensional transport
systems~\cite{appert-rolland_universal_2008,lecomte_imparato_current_2010,gorissen_finite_2011}
or the geometrical features of the system
configurations in glassy systems~\cite{bodineau_toninelli_2011,pitard_dynamic_2011}.
From a broader point of view, other quantities than the activity (such
as the time and space integrated current) present a large deviation
function which becomes singular in the infinite size
limit~\cite{lecomte_imparato_current_2010,bodineau_distribution_2005,bertini_non_2006},
also describing a dynamical phase transition.

\section{Model and description of the coexistence of active and inactive regions}
\label{sec:interface}

\subsection{Large deviations of the activity in the Fredrickson-Andersen model}
\label{subsec:ldfFA}

We focus on a one-dimensional version of the
Fredrickson-Andersen~\cite{fredrickson_kinetic_1984} model (FA model),
in periodic boundary conditions.  It consists of a lattice of size $L$
described by occupation numbers $\mathbf n= (n_i)_{1\leq i\leq L}$ with sites $0$
and $L$ identified. Each site $i$ is either occupied (or `active',
$n_i=1$) or empty (or `inactive', $n_i=0$).  Transition rates are
\begin{align}
  W\big(n_i=0 \to n_i=1\big) &= c\, C_i
  \label{eq:ratesFA1}
\\
  W\big(n_i=1 \to n_i=0\big) &= (1-c) C_i
  \label{eq:ratesFA2}
\end{align}
with  $ C_i=n_{i-1}+n_{i+1}$. The kinetic constraint $C_i$ encodes the ``dynamical facilitation''
rule: active regions favour activity in their vicinity.  Compared to
the unconstrained system ($C_i=1$), the kinetic constraint does not
modify the steady equilibrium state: each site has a Poissonian
occupation number of density $c$, excepted that the configuration where all
sites are inactive is dismissed.
The kinetic constraint however modifies the dynamical relaxation of
correlation functions~\cite{fredrickson_kinetic_1984} with features
similar to those of glassy systems.

It has been shown that the dynamical free energy $\psi_L(s)$ of a system
of size $L$, defined as
\begin{equation}
  \big\langle e^{-sK_t}\big\rangle \underset{t\to\infty}\sim e^{t\psi_L(s)}
  \label{eq:defpsiL}
\end{equation}
 presents a
first order transition in the infinite size limit~\cite{garrahan_dynamical_2007,garrahan_first-order_2009}:
\begin{equation}
  \frac 1L \psi_L(s) \underset{L\to\infty}{\to} 
  \begin{cases}
  >0 & \text{ if } s\leq 0
\\
  0  &  \text{ if } s> 0
  \end{cases}
  \label{eq:res_phiL_infiniteL}
\end{equation}
Note that the mean activity $\frac 1t \langle K_t\rangle_s = \frac {\langle K_t
  e^{-sK_t}\rangle}{t\langle e^{-sK_t}\rangle}$, is also $\frac 1t \langle
K_t\rangle_s=-\psi_L'(s)$. 
{The transition can be interpreted as follows:}
\begin{itemize}
\item $s<0$ corresponds to histories where the mean activity $\langle
  K_t\rangle_s$ is larger than the typical one. For these histories, the 
  number of active sites remains extensive with the system size.
\item $s>0$ corresponds to histories where the mean activity $\langle
  K_t\rangle_s$ is smaller than the typical one. For these histories,
  the number of active sites becomes finite in the large size limit.
  In particular for infinite $s$ the value of the large deviation function is
  given by (the opposite of) the escape rate with only one active site in the system~\cite{garrahan_first-order_2009}:
  \begin{equation}
  \lim_{s\to\infty} \lim_{L\to\infty} \psi_L(s) = -r_{\infty}\equiv -2c 
\end{equation}
\end{itemize}

\subsection{Finite-size scaling and an interface model}
\label{subsec:interface}

\begin{figure}[htpb]
  \centering
    \def\svgwidth{.6\columnwidth}

\begingroup
  \makeatletter
  \providecommand\color[2][]{    \errmessage{(Inkscape) Color is used for the text in Inkscape, but the package 'color.sty' is not loaded}
    \renewcommand\color[2][]{}  }
  \providecommand\transparent[1]{    \errmessage{(Inkscape) Transparency is used (non-zero) for the text in Inkscape, but the package 'transparent.sty' is not loaded}
    \renewcommand\transparent[1]{}  }
  \providecommand\rotatebox[2]{#2}
  \ifx\svgwidth\undefined
    \setlength{\unitlength}{350.84444473pt}
  \else
    \setlength{\unitlength}{\svgwidth}
  \fi
  \global\let\svgwidth\undefined
  \makeatother
  \begin{picture}(1,0.73919027)    \put(0,0){\includegraphics[width=\unitlength]{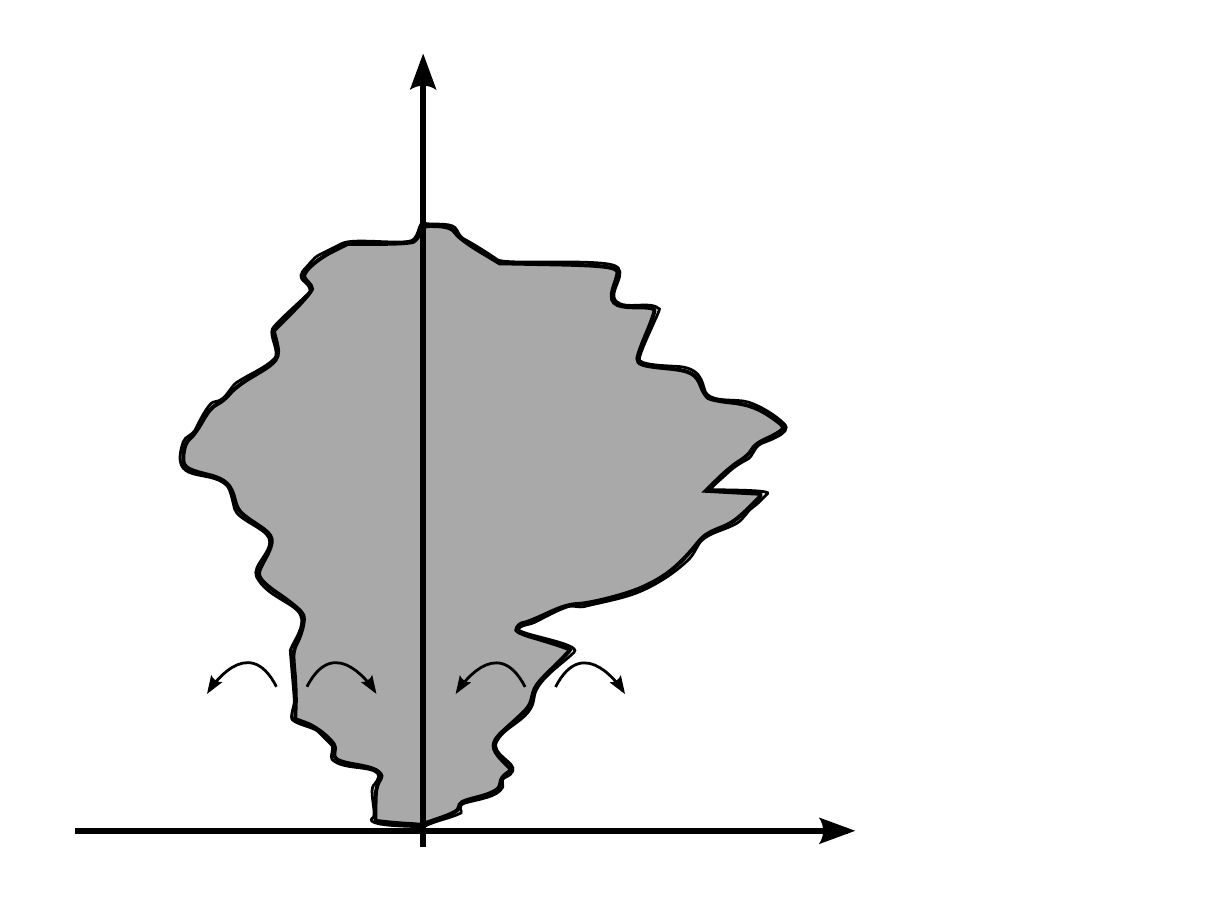}}    \put(0.66018477,0.0034568){\color[rgb]{0,0,0}\makebox(0,0)[lb]{\smash{$x$}}}    \put(0.66697976,0.37929972){\color[rgb]{0,0,0}\makebox(0,0)[lb]{\smash{$x_+(\tau)$}}}    \put(0.36403726,0.69852952){\color[rgb]{0,0,0}\makebox(0,0)[lb]{\smash{$\tau$}}}    \put(-0.00497998,0.37089686){\color[rgb]{0,0,0}\makebox(0,0)[lb]{\smash{$x_-(\tau)$}}}    \put(0.31424981,0.55982872){\color[rgb]{0,0,0}\makebox(0,0)[lb]{\smash{$t$}}}    \put(0.36962639,0.37415428){\color[rgb]{0,0,0}\makebox(0,0)[lb]{\smash{Active}}}    \put(0.18150883,0.21221992){\color[rgb]{0,0,0}\makebox(0,0)[lb]{\smash{$q$
}}}    \put(0.2682281,0.21221992){\color[rgb]{0,0,0}\makebox(0,0)[lb]{\smash{$p$
}}}    \put(0.38560506,0.21204526){\color[rgb]{0,0,0}\makebox(0,0)[lb]{\smash{$p$
}}}    \put(0.47232434,0.21204526){\color[rgb]{0,0,0}\makebox(0,0)[lb]{\smash{$q$
}}}    \put(0.3288568,0.00306461){\color[rgb]{0,0,0}\makebox(0,0)[lb]{\smash{$0$}}}    \put(0.37693729,0.32945952){\color[rgb]{0,0,0}\makebox(0,0)[lb]{\smash{region}}}  \end{picture}\endgroup
  \caption{Model for the space-time configuration of the system in the interfacial regime $\lambda>\lambda_\cc$.
  An island of activity density $\mk=4 c^2(1-c)$ is delimited by two non-crossing biased random walks $x_+(\tau)$
and $x_-(\tau)$, constrained to start at $0$ and end at $0$ at time $t$.}
  \label{fig:interfaces} 
\end{figure}
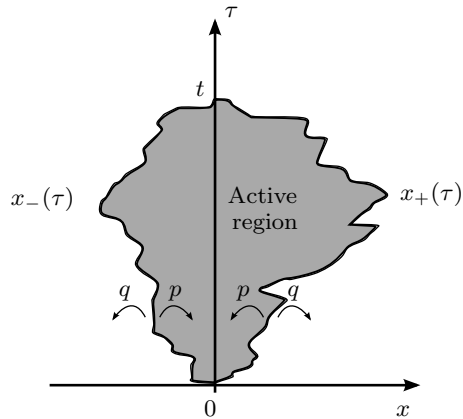

In~\cite{bodineau_toninelli_2011} two of the authors have considered a
different scaling regime by focusing on values of $s$ of the order
$s=\lambda/L$ and for the function
\begin{equation}
  \phi_L(\lambda) \equiv \psi_L(\lambda/L)
  \label{eq:defphiL}
\end{equation}
they conjectured that there exists a critical value $\lambda_\cc>0$ of $\lambda$ such that
\begin{equation}
  \phi(\lambda)=\lim_{L\to\infty} \phi_L(\lambda) =
  \begin{cases}
    -\mk\lambda & \text{ for } \lambda\leq \lambda_\cc \\
    -\Sigma & \text{ for } \lambda\geq \lambda_\cc
  \end{cases}
\label{eq:result_phiL_infiniteL}
\end{equation}
where $\mk = \frac 1{Lt} \langle K_t\rangle = 4 c^2(1-c)$ is the mean
average activity in the system, and $\Sigma$ is the surface
tension accounting for the cost of maintaining an interface between an
active and an inactive region in the system for a long time.
In~\cite{bodineau_toninelli_2011}, the limit~\eqref{eq:result_phiL_infiniteL} was derived only in a range of values $\lambda < \lambda_0$ and $\lambda > \lambda_1$ for some parameters $0< \lambda_0 < \lambda_1$ and not up to the conjectured critical value $\lambda_\cc =  \frac{\Sigma}{\mk}$.

Note that $\Sigma\neq r_\infty$: the typical configurations of the system
at finite $\lambda>\lambda_\cc$ are not given by those of the $s\to\infty$ limit.
In particular, they present more than a finite number of active sites.
Our aim in this article is to identify the typical configurations
occurring at $\lambda>\lambda_\cc$, and to determine the finite size
corrections that they imply on the infinite size
result~\eqref{eq:result_phiL_infiniteL}.
These configurations are interesting to characterise because they are the first to appear
when increasing $\lambda$ (that is, they are the first to appear when
considering histories of the system displaying an activity $K_t$ lower than the typical one).

\medskip

We introduce now a simplified dynamics in order to model the
configurations at $\lambda>\lambda_\cc$.  In the slow activity regime,
the system can be described at a macroscopic level by a small active
``island'' of mean activity $\mk$ in a large sea of an inactive region
(see figure~\ref{fig:interfaces}).
In the unbiased dynamics ($\lambda=0$), the inactive region would be
invaded and become active. Thus at the macroscopic level, an interface
between an active and an inactive region should perform a biased
random walk with effective jump rates $p,q$ which take into account
the growth of the active region.  When $\lambda>\lambda_\cc$, the
growth of the active region is penalised as the activity of the system
is proportional to the area of the active droplet.

More precisely, the boundaries $x_+(t)$ and $x_-(t)$ of the active
region perform non-crossing random walks of jump rate $p$ (resp. $q$)
to the left (resp. right) for $x_+(t)$ and mirror rates for
$x_-(t)$. For simplicity the walks are constrained to start from $x=0$
at time $0$, and to come back to $0$ at final time~$t$ (this
assumption does not change the large time asymptotics).
In this effective description, the total activity in the system is
proportional to the area of the active droplet and approximated by
$\mk\int_0^t d\tau\:[x_+(\tau)-x_-(\tau)]$ with $\mk = 4 c^2(1-c)$ the
mean density of activity.  Thus the counterpart of $ \left\langle
  e^{-sK_t}\right\rangle$ reads
\begin{equation}
 Z_{\text{eff}}(s,t)\equiv 
  \frac{\left\langle e^{-s\mk\int_0^t d\tau\:[x_+(\tau)-x_-(\tau)]}\ \delta(x_\pm(t)=0)\right\rangle_{p,q}}
   {\big\langle \delta_\pm(x(t)=0)\big\rangle_{p,q}}
\end{equation}
where 
$\left\langle \cdot \right\rangle_{p,q}$ denotes the average over
trajectories $x_\pm(\tau)_{0\leq\tau\leq t}$ without constraint at
final time.

This is the simplest model one can think of  to represent the
separation between active and inactive regions in the system. In
particular adding more interfaces would lead to a metastable
situation where the active regions eventually merge together  to
form a unique island of activity.
We think that the interface model represents the correct dynamics of
the system at large scale, but we have not found a rigorous derivation
starting from the microscopic dynamics. However, the numerical results
of section~\ref{sec:numerical_results} support the scaling derived
from the simplified model~\eqref{eq:result_philambda_interface}.

Thus we conjecture that the finite size corrections to 
the large deviation function $\phi_L(\lambda)$~\eqref{eq:defphiL} for $\lambda>\lambda_\cc$  are related to 
\begin{equation}
\hat  \phi_L(\lambda) = \lim_{t \to \infty} \; \frac{1}{t} \log Z_{\text{eff}}(\tfrac{\lambda}{L},t)
\label{eq:defphihat}
\end{equation}
Inspired by the study of interfaces in the static Ising model~\cite{hryniv_universality_2004,velenik_entropic_2004},
and using results from Brownian bridge theory~\cite{majumdar_airy_2005},
we show in appendix~\ref{app:scaling_interface} that this leads to the
following scaling at large $L$
\begin{equation}
\hat  \phi_L(\lambda) =  - 4\sqrt{pq} \left(\frac{\lambda\mk}{4L\sqrt{pq}}\right)^{\frac 23} 2^{-\frac 13}\alpha_1
  \label{eq:result_philambda_interface 0}
\end{equation}
where $\alpha_1\approx 2.3381...$ is the first zero of the Airy
function on the negative real axis.
As a consequence, we expect that the finite size scaling of the microscopic model should be given by~\eqref{eq:result_philambda_interface 0} plus the extra cost $-\Sigma$ for creating the interfaces
\begin{equation}
  \phi_L(\lambda) = -\Sigma - 4\sqrt{pq} \left(\frac{\lambda\mk}{4L\sqrt{pq}}\right)^{\frac 23} 2^{-\frac 13}\alpha_1
  \label{eq:result_philambda_interface}
\end{equation}
for appropriate choice of the effective parameters $p,q$ (see section~\ref{subsec: effective} for a discussion on the effective jump rates).
In other words the interface model we have considered leads to
$L^{-\frac 23}$ corrections to the constant $-\Sigma$.

\section{Numerical results}
\label{sec:numerical_results}

\begin{figure}[t]
  \centering
  \setlength{\unitlength}{.66\columnwidth}
  \begin{picture}(1,0.59645782)
    \put(0,0){\includegraphics[width=\unitlength]{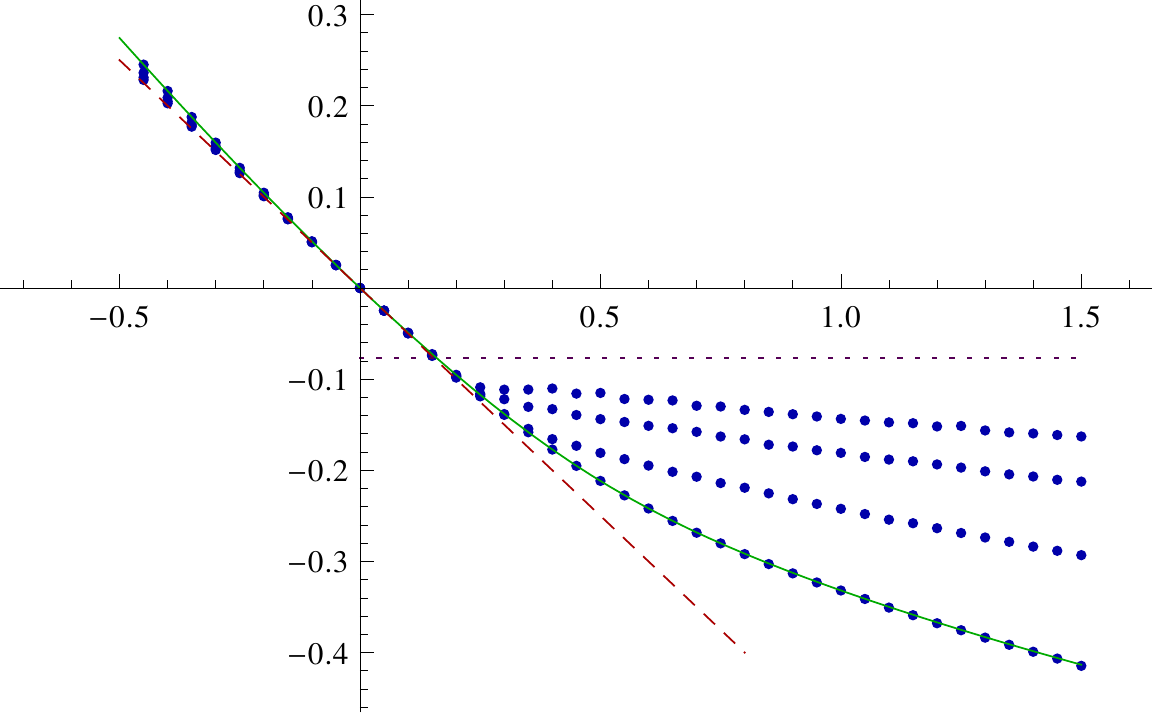}}
    \put(0.35,0.6){\color[rgb]{0,0,0}\makebox(0,0)[lb]{\smash{$\phi_L(\lambda)$}}}
    \put(.98,0.4){\color[rgb]{0,0,0}\makebox(0,0)[lb]{\smash{$\lambda$}}}
  \end{picture}
  \caption{Evaluation of the large deviation function
    $\phi_L(\lambda)$ using the cloning algorithm (blue circles,
    increasing sizes $L\in\{8,16,32,64\}$ from bottom to top at
    positive $\lambda$), and using direct diagonalisation of the
    operator of evolution~\eqref{eq:defWofs_generic} (plain green line,
    $L=8$ run as a check).  The red dashed line is the infinite $L$
    result $-\mk\lambda$ for $\lambda<\lambda_\cc$, while the purple
    dotted horizontal line is the infinite $L$ result $-\Sigma$ for
    $\lambda>\lambda_\cc$.  We took $c=\frac 12$.  }
  \label{fig:psi-of-lambda_FA}
\end{figure}

\subsection{Results from the cloning algorithm (1): the free energy}
\label{sec:numerical_results_cloning_phi}

To investigate whether the finite size
corrections~\eqref{eq:result_philambda_interface} inferred from the
interface model are correct, we have measured $\phi_L(\lambda)$ in
numerical simulations.  Since large deviations are by definition
difficult to measure, a direct sampling of $\psi_L(s)$
through~\eqref{eq:defpsiL} is not achievable. We have resorted to a
continuous-time
version~\cite{lecomte_numerical_2007,tailleur_simulation_2009} of the
Giardin\`a-Kurchan-Peliti cloning
algorithm~\cite{giardina_direct_2006} in which the dynamics is
modified so as to make the large deviation typical, at the price of
mutation/selection rules between a large number of copies of the
system (see~\cite{giardina_simulating_2011} for a review on cloning
algorithms).  Those algorithms have already been used to determine
large deviation functions in lattice
gases~\cite{garrahan_dynamical_2007,hurtado_current_2009,turci_large_2011}
but not in the scaling regime $s=\lambda/L$ that we consider in this
article.

A first result (figure~\ref{fig:psi-of-lambda_FA})
is that the large deviation function agrees qualitatively with
the conjectured infinite size result~\eqref{eq:result_phiL_infiniteL}:
the large deviation function $\phi_L(\lambda)$ tends to become
linear for $\lambda<\lambda_\cc$ and constant
for $\lambda>\lambda_\cc$ as $L$ increases.
The critical value is determined as $\lambda_\cc = \Sigma / \mk$.

\begin{figure}[t]
  \centering
      \setlength{\unitlength}{.49\columnwidth}
  \begin{picture}(1,0.59645782)
    \put(0,0){\includegraphics[width=\unitlength]{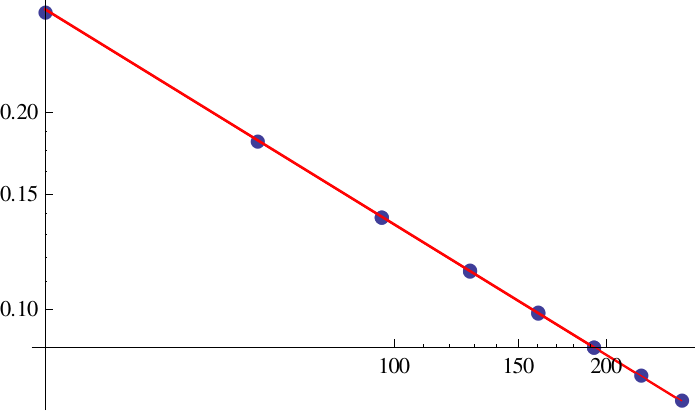}}
    \put(0.15,0.55){\color[rgb]{0,0,0}\makebox(0,0)[lb]{\smash{$\phi_L(\lambda_0)+\Sigma$}}}
    \put(.94,0.12){\color[rgb]{0,0,0}\makebox(0,0)[lb]{\smash{$L$}}}
  \end{picture}
  \begin{picture}(1,0.59645782)
    \put(0,0){\includegraphics[width=\unitlength]{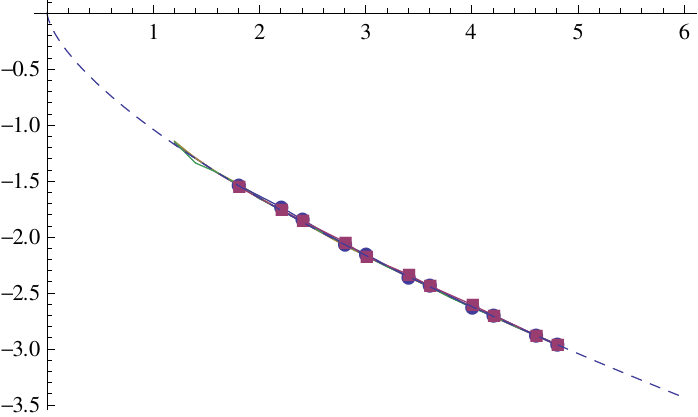}}
    \put(0.1,0.05){\color[rgb]{0,0,0}\makebox(0,0)[lb]{\smash{$L^\alpha\big(\phi_L(\lambda)+\Sigma\big)$}}}
    \put(.96,0.48){\color[rgb]{0,0,0}\makebox(0,0)[lb]{\smash{$\lambda$}}}
  \end{picture}
  \caption{\textbf{(Left)} Log-log plot of $\phi_L(\lambda_0)+\Sigma$, for $\Sigma=0.077$, at fixed $\lambda_0=4.6$ as a function of~$L$: 
    the numerical evaluation (blue dots) fits with a power law corresponding to the exponent exponent $\alpha=\frac 23$ (red line).
    \textbf{(Right) } Plot of $L^\alpha(\phi_L(\lambda)+\Sigma)$ for different values of $L$ ($L\in\{64, 96, 128, 160, 192, 256, 320\}$).
   The curves collapse on a single master curve $-1.05\lambda^\alpha$ (dashed blue),
   in agreement with the interfacial model result~\eqref{eq:result_philambda_interface}.
   The parameter $c$ of the model is $c=\frac 12$.
}
  \label{fig:psiclones_scalings}
\end{figure}
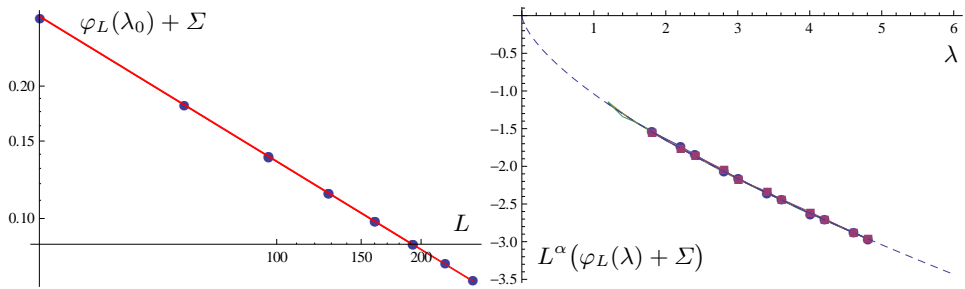

The scaling of the deviations from the infinite size result is
examined in figure~\ref{fig:psiclones_scalings}.  In agreement with
the interfacial model result~\eqref{eq:result_philambda_interface},
$\phi_L(\lambda)+\Sigma$ scales in $L^{-\alpha}$ at fixed $\lambda$
(figure~\ref{fig:psiclones_scalings}, left), with $\alpha=\frac 23$,
while the rescaled curves $L^\alpha(\phi_L(\lambda)+\Sigma)$ collapse
onto a master curve $-A_1\lambda^\alpha$, with $A_1\simeq 1.04$
(figure~\ref{fig:psiclones_scalings}, right).
Simulations were performed at mean density $c=\frac 12$, but the
results and the scaling analysis we present do not depend on this
value.

\subsection{Results from the cloning algorithm (2): the density}
\label{sec:numerical_results_cloning_density}

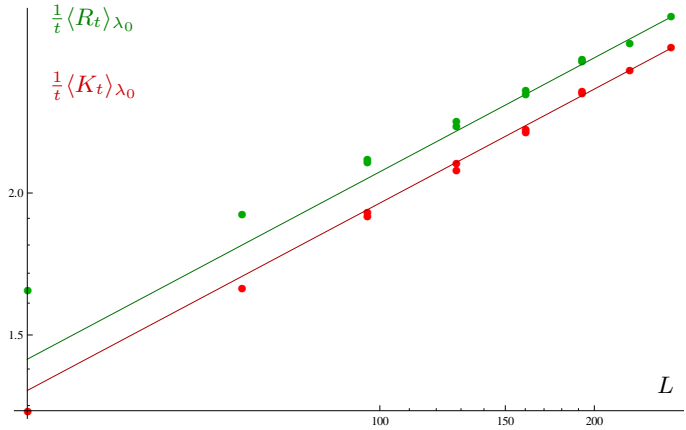
\begin{figure}[htpb]
  \centering
    \setlength{\unitlength}{.7\columnwidth}
  \begin{picture}(1,0.59645782)
    \put(0,0){\includegraphics[width=\unitlength]{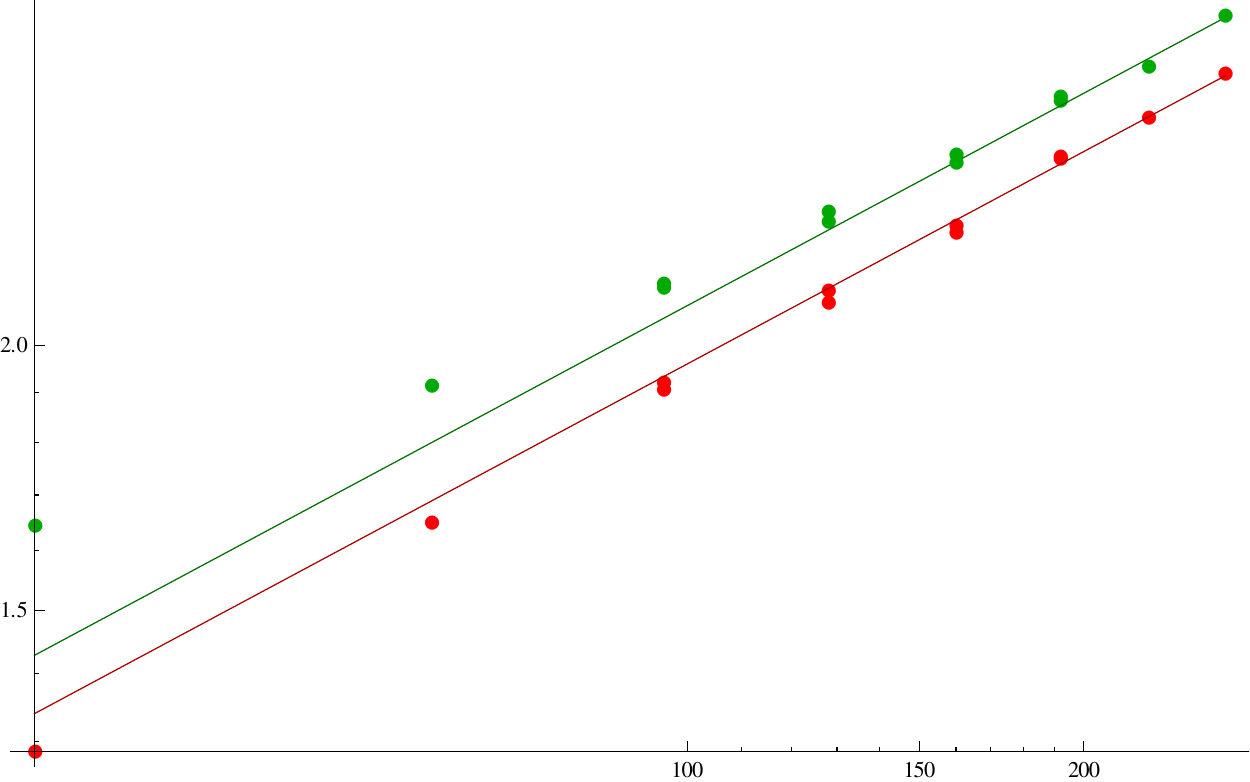}}
    \put(0.06,0.6){\color[rgb]{0,.5,0}\makebox(0,0)[lb]{\smash{$\frac 1t \langle R_t\rangle_{\lambda_0}$}}}
    \put(0.06,0.5){\color[rgb]{.8,0,0}\makebox(0,0)[lb]{\smash{$\frac 1t \langle K_t\rangle_{\lambda_0}$}}}
    \put(.96,0.05){\color[rgb]{0,0,0}\makebox(0,0)[lb]{\smash{$L$}}}
  \end{picture}
  \caption{In red (resp. green): log-log plot of $\frac 1t \langle
    K_t\rangle_{\lambda_0}$ (resp. $\frac 1t \langle R_t\rangle_{\lambda_0}$) at
    fixed $\lambda_0=4.6$ as a function of~$L$ (same values as in
    figure~\ref{fig:psiclones_scalings}). Dots are numerical data
    while lines are power law asymptotics with exponent $1-\alpha=\frac 13$.
  Duplicated dots correspond to different runs.}
  \label{fig:meanKclones}
\end{figure}

One may test another consequence of the interfacial model by computing the density of active sites.
Taking the derivative with respect to $s$ in~\eqref{fig:psiclones_scalings} leads to 
$\frac 1t \langle K_t\rangle_s=- \psi_L'(s)$.
Thus using the relation~\eqref{eq:defphiL} implies that  $\frac 1t \langle
K_t\rangle_\lambda=-L \phi_L'(\lambda)$ and from~\eqref{eq:result_philambda_interface} the mean activity $\frac 1t
\langle K_t\rangle_\lambda$ for histories weighted by $e^{-\lambda K_t/L}$
scales as $L^{\frac 13}$.
Therefore, one expects that the average width of the active droplet is $L^{\frac 13}$ when $\lambda>\lambda_c$.
Strong finite-size effects are still present (figure~\ref{fig:meanKclones})
and wouldn't allow to check precisely the power $\frac 23$ of the
scaling relation~\eqref{eq:result_philambda_interface}.
To understand the origin of these corrections a useful tool is the
escape rate $r(\C)$ from a configuration $\C=(n_i)_{1\leq i\leq L}$
defined as the sum of the jump rates from $\C$: $r(\C)=\sum_{i=1}^L
\big[(1-c)n_i+c(1-n_i)\big](n_{i-1}+n_{i+1})$.  As shown in
appendix~\ref{app:identity_ldfKR}, the fluctuations of $K_t$ and of the
time integral of the escape rate $R_t=\int_0^td\tau\:r(\C(\tau))$ are
closely related (this result is valid in general):

\begin{equation}
 { \psi_L(s) = \frac 1t \langle K_t\rangle_s - \frac 1t \langle R_t\rangle_s }
 \label{eq:symmetrypsiKRtext}
\end{equation}

 In the interfacial regime $\lambda>\lambda_\cc$ , $\phi_L(\lambda)$ goes to a
constant $-\Sigma$ when $L$ goes to infinity, while both $\frac 1t
\langle K_t\rangle_\lambda$ and $\frac 1t \langle R_t\rangle_\lambda$ grow
with $L$ so that~\eqref{eq:symmetrypsiKRtext} describes the cancellation
between those growths.
More precisely, assuming the scaling form
\begin{align}
  \phi_L(\lambda) &= -\Sigma + L^{-\alpha} \phi^1(\lambda) + o(L^{-\alpha})
  \label{eq:finitesize_scalingform_psi_text}
\end{align}
one sees by differentiating~\eqref{eq:symmetrypsiKRtext} that both $
\langle K_t\rangle_\lambda$ and $ \langle R_t\rangle_\lambda$ scale in the
same way with $L$
\begin{align}
  \frac 1t \langle K_t\rangle_\lambda \sim L^{1-\alpha}\: k_1(\lambda) + o(L^{1-\alpha})
\qquad
  \frac 1t \langle R_t\rangle_\lambda \sim L^{1-\alpha}\: r_1(\lambda) + o(L^{1-\alpha})
  \label{eq:KtRt}
\end{align}
where the exponent $1-\alpha$ comes from the relation $\frac 1t \langle K_t\rangle_\lambda=-L \phi'_L(\lambda)$
and $r_1=k_1$.

We also note that at $c=\frac 12$, $r(\C)$ is (twice) the total number
of active sites so that the $\frac 1t \langle R_t\rangle_\lambda$ also
represents the mean density of active sites at fixed $\lambda$,
which thus scales as $L^{1-\alpha}=L^{\frac 13}$ (we
expect the result to hold also for $c\neq\frac 12$ although $\frac 1t \langle R_t\rangle_\lambda$ is not the mean density anymore).
We conclude that a numerical check or evaluation of the exponent $\alpha$
is better done on $\phi_L(\lambda)$ than
on the mean activity, the density or the escape rate, since those quantities
present large finite-size corrections 
\emph{e.g.} of the form 
\begin{align}
  \frac 1t \langle K_t\rangle_\lambda &\sim 
  L^{1-\alpha}\: k_1(\lambda) 
  + L^\gamma k_2(\lambda) + C_{K}+ O(L^{-\alpha})
\\ 
  \frac 1t \langle R_t\rangle_\lambda &\sim L^{1-\alpha}\: r_1(\lambda) 
  + L^\gamma r_2(\lambda) + C_{R}+ O(L^{-\alpha})
\end{align}
where by cancellation from~\eqref{eq:symmetrypsiKRtext} $k_1=r_1$, $k_2=r_2$ and $C_K-C_R=-\Sigma$.

\subsection{Results in an independent site approximation}
\label{sec:numerical_results_Bernoulli}

We now consider another approach to compute the dynamical free energy
$\phi_L(\lambda)$.  It can be shown that $\phi_L(\lambda)$ is the
largest eigenvalue of a symmetric operator of evolution~$\mathbb
W^\sym_\lambda$~\cite{lecomte_thermodynamic_2007} (see also
appendices~\ref{app:identity_ldfKR} and~\ref{app:Bernoulli}) which
acts on the vector space of all configurations $\{|\mathbf n\rangle\}$ of the system. We thus
have the Courant-Fisher equality
\begin{equation}
  \phi_L(\lambda) = \max_{|X\rangle\neq 0 } \frac{\langle X|\mathbb W^\sym_\lambda|X\rangle}{\langle X|X\rangle}
  \label{eq:phiL_optim}
\end{equation}
where $|X\rangle=\sum_{\mathbf n} X(\mathbf n)|\mathbf n\rangle$ is a vector.

Restricting the optimisation on laws $X(\mathbf n)$ representing products of Bernoulli distributions 
for evaluating the largest eigenvalue of $\mathbb W^\sym_s$,
one obtains in appendix~\ref{app:Bernoulli} the following estimation for the large deviation
function:  $\phi_L(\lambda)\leq \phi_L^{\text{Bern}}(\lambda)$ with
\begin{equation}
  \phi_L^{\text{Bern}}(\lambda)=
  \max_{\{\rho_i\}}
  \frac
  {\sum_{1\leq i\leq L} \big[e^{-\lambda L}\sqrt{c(1-c)\rho_i(1-\rho_i)}-c(1-\rho_i)-(1-c)\rho_i\big](\rho_{i+1}+\rho_{i-1}) }
  {1-\prod_{1\leq i\leq L}(1-\rho_i)}
\label{eq:psiBern}
\end{equation}
where $0<\sqrt{\rho_i}<1$ is the parameter of the Bernoulli law at site $i$.

\begin{figure}[tpb]
  \centering
    \setlength{\unitlength}{.66\columnwidth}
  \begin{picture}(1,0.59645782)
    \put(0,0){\includegraphics[width=\unitlength]{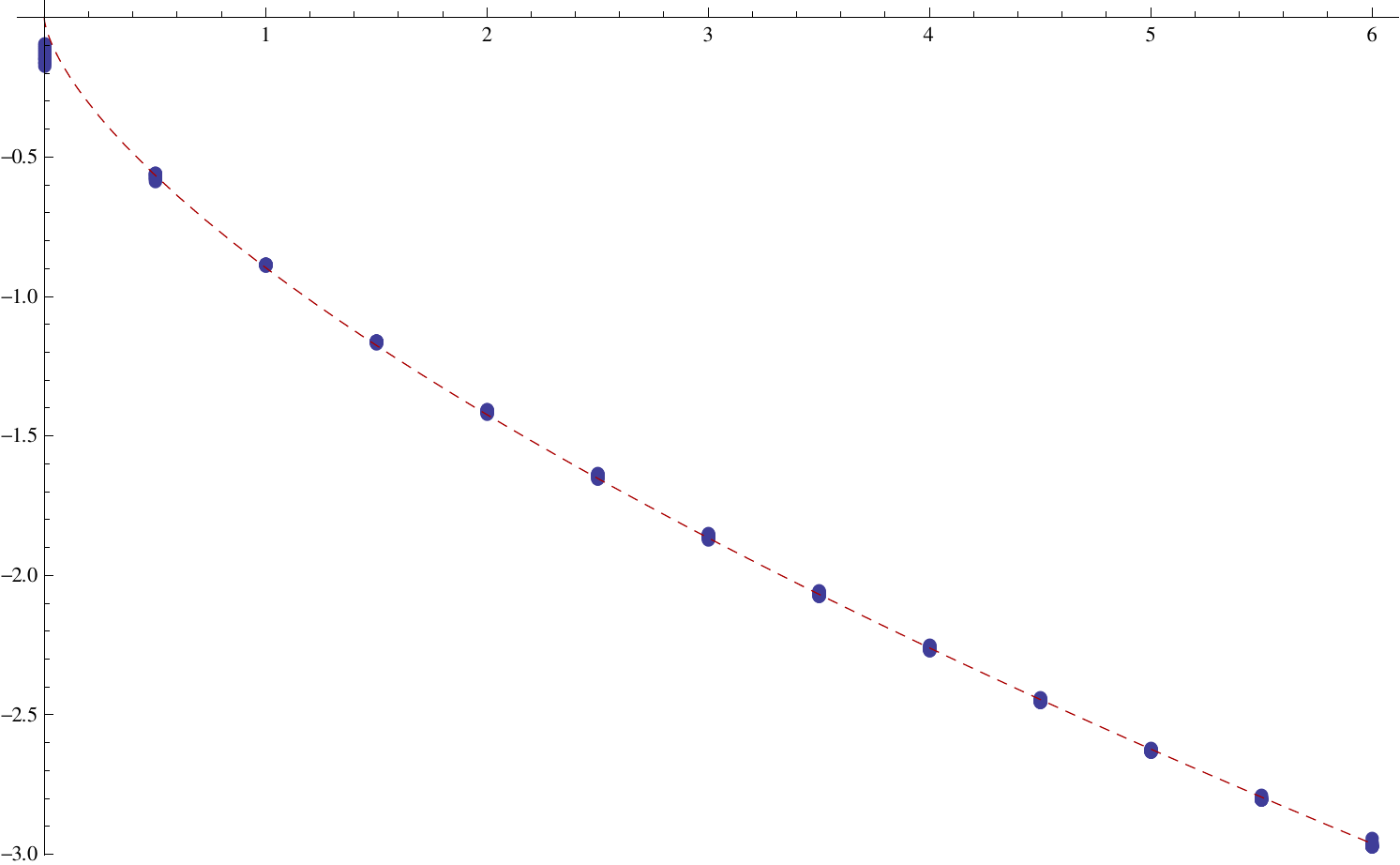}}
    \put(0.05,0.05){\color[rgb]{0,0,0}\makebox(0,0)[lb]{\smash{$L^\alpha(\phi_L(\lambda,L)+\Sigma)$}}}
    \put(.96,0.55){\color[rgb]{0,0,0}\makebox(0,0)[lb]{\smash{$\lambda$}}}
  \end{picture}
  \caption{ Plot of the scaling function
    $L^\alpha(\phi_L(\lambda,L)+\Sigma)$ for different values of $L$
    ($L\in\{32, 36,40,\ldots 96, 100\}$) determined from the Bernoulli
    optimisation~\eqref{eq:psiBern} (blue dots).  The curves collapse
    on a single master curve $-A_1\lambda^\alpha$ (dashed red).  The
    values of $\Sigma$ and $A_1$ are quite different from those of the original FA
    model: here, $\Sigma\approx 0.251$ and $A_1\approx 0.897$.  }
  \label{fig:psiBern} 
\end{figure}

The mean value of the mean fraction of active sites $\frac
1{Lt}\int_0^td\tau\sum_in_i(\tau)$ for histories weighted by $e^{-sK_t}$
is given by
\begin{equation}
  \bar\rho(\lambda,L)\equiv \lim_{t\to\infty} \Big\langle e^{-sK_t}\frac 1{Lt}\int_0^td\tau\sum_in_i(\tau) \Big\rangle / \langle e^{-sK_t}\rangle
  =\langle L|\frac 1L\sum_i\hat n_i|R\rangle
\end{equation}
where $|L\rangle$ and $|R\rangle$ are the left and right eigenvectors of $ \mathbb W_s $ associated to the eigenvalue $\phi_L(s)$.
The corresponding value in the Bernoulli projection reads
\begin{equation}
  \bar\rho^{\text{Bern}}(\lambda,L)\ =\  \frac 1L\:
  \frac
  1
  {1-\prod_{1\leq i\leq L}(1-\rho_i)}
  \sum_{1\leq i\leq L}\rho_i
\end{equation}
for a set ${\{\rho_i\}}$ which optimises~\eqref{eq:psiBern}.
Numerically solving~\eqref{eq:psiBern} one obtains
$\bar\rho^{\text{Bern}}(\lambda,L)$ and the large deviation function. 
The collapse for $\phi^{\text{Bern}}_L(\lambda)$ works again with $\alpha=\frac 23$ (see figure~\ref{fig:psiBern}),
while again $\bar\rho^{\text{Bern}}(\lambda,L)$ displays stronger finite-size effects.

We expect the true minimiser to be very different from the
approximation by a Bernoulli distribution.  Indeed the sites located
at the interface should be very correlated. This explains why the
values of $\Sigma$ and $A_1$ measured within the Bernoulli approximation are
different from those of the original FA model (see
section~\ref{subsec:universality}).

\section{Discussion}
\label{sec:discussion}

\subsection{Determination of the effective jump rates $p$ and $q$, and the surface tension $\Sigma$}
\label{subsec: effective}

The exponent $\alpha=\frac 23$ observed numerically matches the one
predicted by the interface model~\eqref{eq:result_philambda_interface}.
One can also compare the value of the coefficient $A_1$ in the scaling 
\begin{equation}
  \phi_L(\lambda) = -\Sigma - A_1 \Big(\frac\lambda L\Big)^\alpha+o(L^{-\alpha})
  \label{eq:generic_scaling_phiL_coex}
\end{equation}
In a very crude approximation, where correlations are neglected, 
one can imagine the interface as a single active site at position $x(t)$ separating a region with
only inactive sites from an active region sampled according to a
Bernoulli measure with parameter $c$. It this case, the parameters
$p$, $q$ of the interface model defined in section~\ref{subsec:interface}
can be estimated as follows. We have $q=c$: the interface grows at rate $c$
by activating a site on the left of $x(t)$. 
On the other hand, the value of $p$ may be estimated to $p = c(1 -
c)$: due to the kinetic constraint the first active site is
inactivated with rate $1 - c$ provided that the site to its right is
active, which occurs with probability $c$.
From~\eqref{eq:result_philambda_interface} and $\mathbb K =
4c^2(1-c)$, this yields
\begin{equation}
  A_1 =
    4\sqrt{pq} \left(\frac{\mk}{4\sqrt{pq}}\right)^{\frac 23} 2^{-\frac 13}\alpha_1
    = \mk^{\frac 56}\alpha_1 
  \label{eq:exprA1pqK}
\end{equation}
and thus $A_1\simeq 1.312$ for $c=\frac 12$. This is close to the value
$A_1\approx 1.05\pm0.01$ observed numerically, the discrepancy arising
in part from correlations between neighbouring sites around the interface.

\medskip 
The value of the dynamical surface tension $\Sigma$ can be compared to
analytical predictions.  In~\cite{bodineau_toninelli_2011} an
expression of $\Sigma$ was derived for the East and the FA models with
fixed boundary condition. A similar expression holds for
the periodic FA model we are interested in:
\begin{equation}
  \Sigma =- \lim_{L\to\infty} \sup_{P} \big\langle \sqrt{P}\big|\mathbb W^\sym(L) \big|\sqrt{P}\big\rangle
  \label{eq:formulaSigma_max}
\end{equation}
where the supremum is performed over all probability distributions on
the set of configurations $\{\mathbf n\}$ of the system, with the
condition that site $1$ is inactive: $n_1=0$.
\begin{figure}[t]
  \centering
    \setlength{\unitlength}{.66\columnwidth}
  \begin{picture}(1,0.59645782)
    \put(0,0){\includegraphics[width=\unitlength]{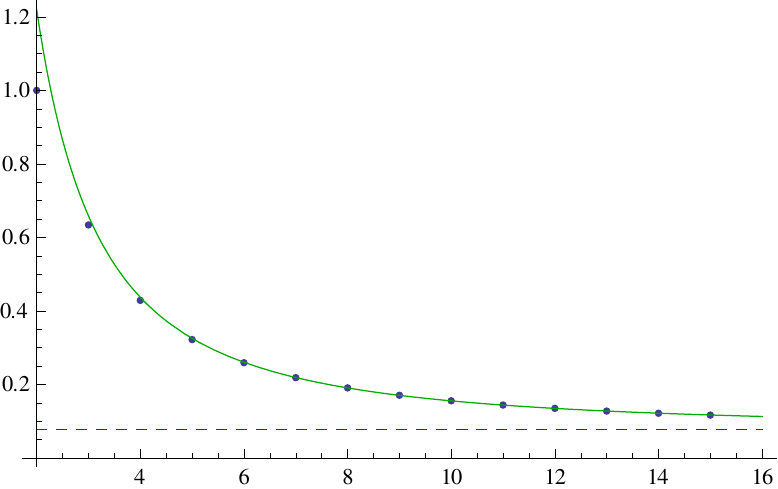}}
    \put(0.08,0.58){\color[rgb]{0,0,0}\makebox(0,0)[lb]{\smash{$\Sigma_L$}}}
    \put(1.05,0.03){\color[rgb]{0,0,0}\makebox(0,0)[lb]{\smash{$L$}}}
  \end{picture}
  \caption{ Plot of the surface tension $\Sigma_L$ at size $L$
    obtained by diagonalisation of the operator appearing
    in~\eqref{eq:Sigma_maxSpPWP} (blue dots). To determine the surface
    tension $\Sigma = \lim_{L\to\infty} \Sigma_L$, we performed the
    fit $\Sigma_L=\Sigma + A L^{-\frac 53}$ (green line).  The
    exponent $\frac 53$ was chosen as the most likely one in regards
    of the numerical results.  The result $\Sigma=0.0771$ (horizontal
    red dashed line) is very close from the one obtained with the
    cloning algorithm ($0.077$, see
    figure~\ref{fig:psiclones_scalings}). }
  \label{fig:SigmaL} 
\end{figure}
%
The vector $|\sqrt{P}\big\rangle$ is the vector of components $\sqrt{P(\mathbf n)}$.
The symmetrized operator of evolution $\mathbb W^\sym(L)$ for a system
of size $L$ is defined in appendix~\ref{app:Bernoulli}.
Note that the formulation of this extremalisation principle
in~\cite{bodineau_toninelli_2011} involves a Dirichlet form which is
equal to the expression maximised in~\eqref{eq:formulaSigma_max}, up to
a sign.
Defining now the projector $\mathbb P_L$ onto the configurations with
site 1 inactive, we can replace $\mathbb W^\sym(L)$ by $\mathbb P_L
\mathbb W^\sym(L) \mathbb P_L$ in~\eqref{eq:formulaSigma_max} and
relax the condition on $P$. This shows that:
\begin{equation}
  \Sigma = \lim_{L\to\infty} \Sigma_L \qquad\text{with}\quad
  \Sigma_L  = - \max \operatorname{Sp} \big (\mathbb P_L
  \mathbb W^\sym(L) \mathbb P_L \big)
  \label{eq:Sigma_maxSpPWP}
\end{equation}
where $\operatorname{Sp}$ denotes the spectrum of an operator.
We thus have expressed the dynamical surface tension using the maximum
eigenvalue of an operator, in a similar way as for the free energy
$\psi_L(s)$ in~\eqref{eq:phiL_optim}. 
The value of $\Sigma_L$ was computed for systems sizes $L\leq 15$ by
direct diagonalisation (figure~\ref{fig:SigmaL}).  By fitting the results using a reasonable form
of the finite-size corrections we obtain the value $\Sigma=0.0771$,
which is numerically compatible to the one obtained from the cloning
algorithm $\Sigma=0.077 \pm 0.0005$. 
Note that the finite-size surface tension $\Sigma_L$ displays strong
finite-size effects fitted through the form $\Sigma_L=\Sigma+ A L^{- \frac 53}$,
but we have no theoretical justification for the power $\frac 53$.

\subsection{Universality}
\label{subsec:universality}

The Bernoulli approximation developed in
part~\ref{sec:numerical_results_Bernoulli} presents the same scaling
exponent $\alpha=\frac 23$ with different constants $\Sigma$ and
$A_1$ (see figure~\ref{fig:psiBern}).
This tells that, on the one hand, the independent site approximation
of the Bernoulli optimisation~\eqref{eq:psiBern} is quite far
from being correct: the large deviations do not match at all.
This fact was also checked by diagonalising the operator of evolution
$\mathbb W_s$ and finding the eigenvector associated to the maximal
eigenvalue $\psi_L(s)$, for small system sizes ($L\leq 15$). We
found that the corresponding state can't be factorised on independent 
sites and presents correlations.
On the other hand, the robustness of the exponent $\frac 23$ 
is an indication that other models in the same class also
present (an) interface(s) in the coexistence regime.

Besides, let us note that the same expression as~\eqref{eq:psiBern}
appears when considering the field theory associated to the operator
of evolution $\mathbb
W_\lambda^\sym$~\cite{lecomte_thermodynamic_2007} and computing
$\phi_L(\lambda)$ assuming a time-independent saddle-point. This tells
that such an assumption is not valid.

\begin{figure}[!t]
  \begin{center}
    \def\svgwidth{.75\columnwidth} 

\begingroup
  \makeatletter
  \providecommand\color[2][]{    \errmessage{(Inkscape) Color is used for the text in Inkscape, but the package 'color.sty' is not loaded}
    \renewcommand\color[2][]{}  }
  \providecommand\transparent[1]{    \errmessage{(Inkscape) Transparency is used (non-zero) for the text in Inkscape, but the package 'transparent.sty' is not loaded}
    \renewcommand\transparent[1]{}  }
  \providecommand\rotatebox[2]{#2}
  \ifx\svgwidth\undefined
    \setlength{\unitlength}{470.07930298pt}
  \else
    \setlength{\unitlength}{\svgwidth}
  \fi
  \global\let\svgwidth\undefined
  \makeatother
  \begin{picture}(1,0.59190239)    \put(0,0){\includegraphics[width=\unitlength]{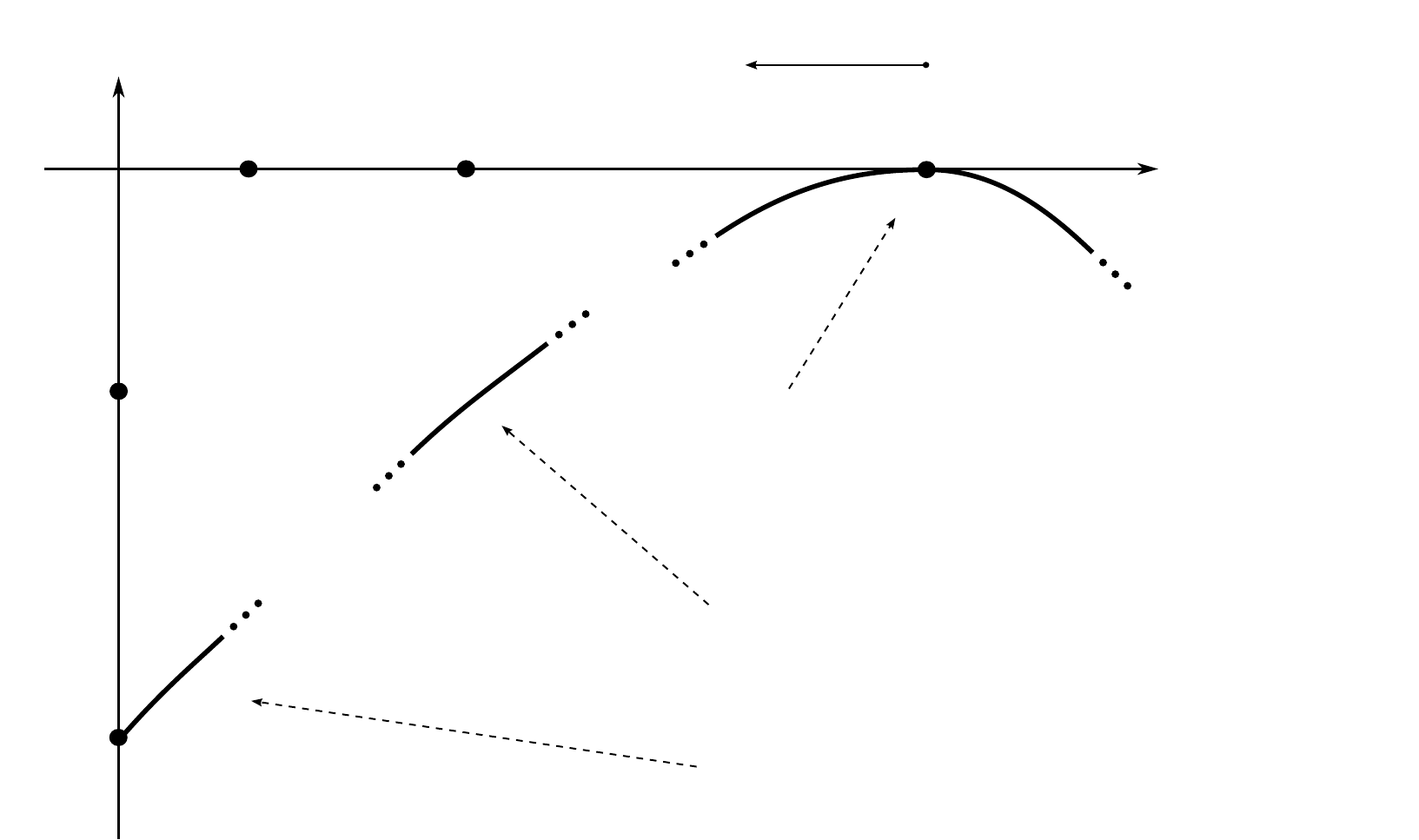}}    \put(0.00772188,0.3078807){\color[rgb]{0,0,0}\makebox(0,0)[lb]{\smash{$-\Sigma$}}}    \put(-0.00315771,0.06160005){\color[rgb]{0,0,0}\makebox(0,0)[lb]{\smash{$-r_\infty$}}}    \put(0.85207096,0.46570641){\color[rgb]{0,0,0}\makebox(0,0)[lb]{\smash{$k$}}}    \put(0.02245412,0.56604239){\color[rgb]{0,0,0}\makebox(0,0)[lb]{\smash{$\pi_L(k)$}}}    \put(0.12816762,0.49430975){\color[rgb]{0,0,0}\makebox(0,0)[lb]{\smash{$L^{-1}$}}}    \put(0.29118249,0.49444697){\color[rgb]{0,0,0}\makebox(0,0)[lb]{\smash{$L^{-\frac 23}$}}}    \put(0.52575821,0.27587802){\color[rgb]{0,0,0}\makebox(0,0)[lb]{\smash{$\pi_L(k) = -L\frac{(k-\mathbb K)^2}{2\mathbb K_2}$}}}    \put(0.52575821,0.16103697){\color[rgb]{0,0,0}\makebox(0,0)[lb]{\smash{$\pi_L(k) = -\Sigma -\frac{4A_1^3}{27L^2} k^{-2}$}}}    \put(0.52575821,0.04681916){\color[rgb]{0,0,0}\makebox(0,0)[lb]{\smash{$\pi_L(k) = -r_\infty + L k\Big(1-\log \frac{Lk}{2c}\Big)$}}}    \put(0.64067633,0.49526436){\color[rgb]{0,0,0}\makebox(0,0)[lb]{\smash{$\mathbb K$}}}    \put(0.53597195,0.56482875){\color[rgb]{0,0,0}\makebox(0,0)[lb]{\smash{$L^{- 1}$}}}  \end{picture}\endgroup
  \end{center}
  \caption{Schematic plot of the finite-size scaling of the large
    deviation function $\pi_L(k)=\lim_{t\to\infty} \log
    \operatorname{Prob }\big[K_t=tLk\big]$, deduced from the
    finite-size scaling of $\phi_L(\lambda)$. The regime of smallest
    $k$ ($k \ll L^{-1}$) corresponds to configurations with finite
    number of active sites.  The intermediate regime ($L^{-\frac 23}\lesssim
    k \ll \mathbb K - L^{-1}$) corresponds to the regime of phase transition,
    where the large deviation function is linear in the infinite size limit
    (cf. equation~\eqref{eq: LD split 2}).
        The picture for an unconstrained dynamics ($C_i=1$
    in~(\ref{eq:ratesFA1}-\ref{eq:ratesFA2})) has a completely
    different large deviation scaling $\pi^\text{unc}_L(k)=L\pi^\text{unc}_1(k)$ with
    inactive histories exponentially much less likely.  }
  \label{fig:pi_of_k} 
\end{figure}
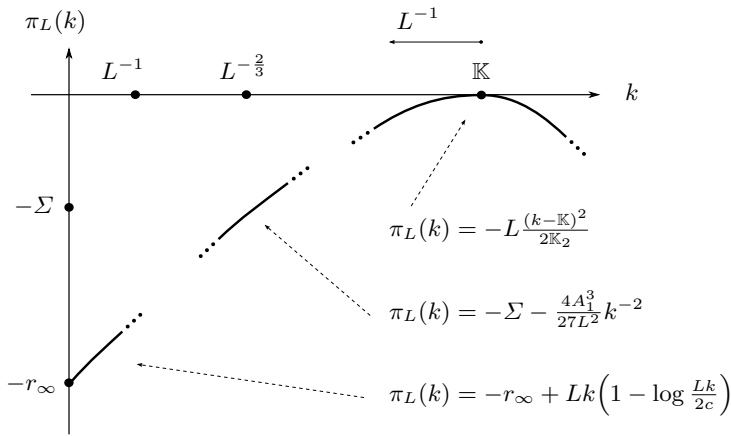

\subsection{Finite size scaling of the large deviation function $\pi_L(k)$}
\label{subsec:pi_L_of_k_finite-size}

At fixed system size $L$, the large deviation function $\pi_L$ defined
as
\begin{equation}
  \text{Prob}[K_t=Lkt] \underset{t\to\infty}\sim e^{t\pi_L(k)}
  \label{eq:defpiL}
\end{equation}
is the Legendre transform of $\psi_L$
\eqref{eq:defpsiL} 
$$
\pi_L(k) = \inf_{s} \left\{ s L k + \psi_L (s)  \right\},
\qquad 
\psi_L( s) = \sup_{k} \left\{ - s L k + \pi_L (k)  \right\}.
$$
Using the parameter $\lambda = s L$, 
 the large deviation function becomes, in the large $L$ limit, for reduced activities
\begin{equation}
\label{eq: LD below}
\forall k\in [0,\mk], \qquad  
\pi(k) = \lim_{L\to\infty} \pi_L(k) = \inf_{\lambda > 0} \left\{ \lambda k + \phi (\lambda)  \right\}
= - \Sigma ( 1- \frac{k}{\mk}) 
\end{equation}
where $\phi$ is defined in~\eqref{eq:result_phiL_infiniteL}. 
This follows from a simple computation
(see~\cite{touchette_large_2009} for a review on ldf in physics and
mathematics).  Note that negative activities cannot be produced.
The physical picture behind the linear behaviour~\eqref{eq: LD below}
is a two step mechanism. An activity deviation of order $k t$ over the
time interval $[0,t]$ is produced by first blocking the system in an
inactive state during a time $( 1- \frac{k}{\mk}) t$ and then letting
the system in the stationary state (with mean activity $\mk$) during a
time $\frac{k}{\mk} t$. This leads to an exponential cost $\Sigma ( 1-
\frac{k}{\mk}) t$. We stress the fact that the switch from the
inactive to the active state occurs on time scales much smaller than
$t$ and therefore has no impact on the large deviation function
obtained in the large $t$ limit.  This behaviour is characteristic of a
first order phase transition between the active and inactive
regime. We examine below the consequences of the finite size
corrections to the large deviations within both regimes.
Results are summarised on figure~\ref{fig:pi_of_k}.

\medskip

\noindent
{\it The active regime:}

For activities larger than $\mk$, the scaling is different as the
constraints do not play a major role and the behaviour is similar to
non-constrained systems, one has
\begin{equation}
\label{eq: LD above}
\forall k > \mk, \qquad  
\hat \pi (k) = 
\lim_{L\to\infty}  \frac{1}{L} \pi_L(k) = \inf_{s < 0} \left\{ s k + \hat\psi (s)  \right\} \, ,
\end{equation}
where $\hat\psi(s)=\lim_{L\to\infty} \frac 1L \psi_L(s)$ which leads to 
\begin{equation}
\forall k > \mk, \qquad  
\text{Prob}[K_t \simeq t L \; k ] \underset{t\to\infty}\sim e^{ t L \hat \pi(k)}
\end{equation}
In particular $\hat \pi$ is expected to be a smooth function for $k>0$
\begin{equation}
\label{eq: LD dvp}
\forall k > \mk, \qquad  
\hat \pi (k) =  -\frac 1{2\mk_2} \big( k - \mk \big)^2 + O \big( (k - \mk)^3 \big)
\end{equation}
where $\mathbb K_2=\frac 1{Lt}\langle K_t^2\rangle_c$ is the variance of the activity.

The finite size scaling asserts that the transition for $\psi_L(s)$
takes place away from $s=0$ and therefore, we expect that for very
small deviations of the activity below $\mk$, $L \hat \pi$ 
approximates the large deviation function $\pi_L (k) = \inf_{\lambda >
  0} \left\{ \lambda k + \phi_L (\lambda) \right\}$.  This would
hold only for $s> \frac{\lambda_c}{L}$, i.e. $k - \mk \approx
\frac{1}{L}$ (by optimising in~\eqref{eq: LD above}). This means that
in a small window around $\mk$ of order $\frac{1}{L}$, the large
deviations are of the form~\eqref{eq: LD dvp}
\begin{equation}
\label{eq: LD dvp 2}
\pi_L (k) =  -\frac{L}{2\mk_2} \big( k - \mk \big)^2 + O (L^{-2})
\end{equation}

\medskip

\noindent
{\it The inactive regime:}

In the inactive regime, the finite size scaling takes into account the deviations of the width of the active droplet which will contribute to the activity deviations when $k$ is close to $0$. 
The finite size Legendre transform~\eqref{eq: LD below} reads
\begin{equation}
\forall k\in [0,\mk], \qquad  
\pi_L(k) = \inf_{\lambda > 0} \left\{ \lambda k + \phi_L (\lambda)  \right\} 
\end{equation}
where the optimal $\lambda$ is given by $k = -\phi_L^\prime (\lambda)$.
The finite size scaling~\eqref{eq:result_philambda_interface} is valid for $\lambda \gg \lambda_\cc$
\begin{equation}
\phi_L (\lambda) = - \Sigma - A_1 \left( \frac{\lambda}{L} \right)^{2/3} 
\qquad \Rightarrow \qquad
k = \frac{2}{3} \frac{A_1}{L^{2/3}} \lambda^{-1/3}\label{eq:1}
\end{equation}
This allows to estimate $\pi_L(k)$ for $k \approx \frac{1}{L^\frac{2}{3}} \lambda^{- \frac{1}{3}}$ and $\lambda \gg 1$
\begin{equation}
\pi_L(k) \simeq -\Sigma -  \frac{4 A_1^3}{27} \;   \frac{1}{L^2 k^2} 
\label{eq:result_philambda_interface pi}
\end{equation}
This scaling should remain correct if the droplet width is very large microscopically, i.e. for $k \gg 1/L$.
For $k \approx {L^{-\frac{2}{3}}} \lambda^{- \frac{1}{3}}$ with $\lambda$ close to zero, the probability of observing an active droplet of width $k L$ should vanish exponentially fast with a rate given by~\eqref{eq:result_philambda_interface pi}. But we will see below that these larger droplets do not contribute to $\pi_L(k)$.

\medskip

\noindent
{\it The intermediate regime:}

We consider now the intermediate deviations in $k$.
The two-step mechanism described earlier has to be slightly modified to take into account the 
finite size corrections found in the active and inactive regimes~\eqref{eq: LD dvp 2}, \eqref{eq:result_philambda_interface pi}.
Thus one expects that  
\begin{eqnarray}
\label{eq: LD split}
&& \text{Prob}[K_t \simeq t L \; k ]  \underset{ t \to \infty} \sim   
\ \\
&& 
\qquad 
\sup_{k_1,k_2 \atop x _1,x_2} 
\left\{  
\exp \left( - t x_1 \;  \frac{L}{2\mk_2} \big( k_1 - \mk \big)^2  
 - t x_2 \;  \Big[ \Sigma +  \frac{4 A_1^3}{27} \;   \frac{1}{L^2 k_2^2} \Big] \right) 
\right\} \nonumber
\end{eqnarray}
where the supremum is taken over the activities $k_1,k_2$ and $x_1, x_2$ such that 
\begin{equation}
\label{eq: LD constraint}
1 = x_1+ x_2, \qquad 
k = x_1 k_1 + x_2 k_2
\end{equation}
Suppose that $k \ll \mk - \frac{1}{L}$.
As $k_1- \mk$ is at most of order $1/L$, one can neglect the deviations with respect to $k_1$ and set $k_1 = \mk$.
This leads to optimise over $k_2$ with
\begin{equation}
k_1 = \mk, \quad 
x_1 = 1-  x_2, \qquad 
x_2 = \frac{\mk - k}{\mk - k_2} \in [0,1]
\end{equation}
Thus~\eqref{eq: LD split} reads
\begin{eqnarray}
\text{Prob}[K_t \simeq t L \; k ]  \underset{t\to\infty}   \sim 
\exp \left( - t (\mk - k) \; \inf_{k_2 \leq k }  
\left\{  G(k_2) \right\}  \right)
\end{eqnarray}
where 
\begin{equation}
G(k_2) =   \frac{1}{\mk - k_2} \;  \big[ \Sigma +  \frac{4 A_1^3}{27} \;   \frac{1}{L^2 k_2^2} \big].
\end{equation}
The function $G$ reaches its minimum for 
\begin{equation}
k^\star = \frac{\left(\sqrt{A^2 L^6 \Sigma^3 \left(A+ \mk^2 L^2 \Sigma \right)}+A \mk L^4 \Sigma^2\right)^{2/3}-A L^2
   \Sigma}{L^2 \Sigma \sqrt[3]{\sqrt{A^2 L^6 \Sigma^3 \left(A+ \mk^2 L^2 \Sigma \right)}+A \mk L^4 \Sigma^2}}
    \underset{ L \to \infty} \sim \left( \frac{2 A}{\Sigma} \right)^{1/3} \ \frac{1}{L^{2/3}} 
\end{equation}
with $A = \frac{4 A_1^3}{27}$. Finally we get
\begin{eqnarray}
\label{eq: LD split 2}
\lim_{t \to \infty} 
 \frac{1}{t} \log 
\text{Prob}[K_t \simeq t L \; k ]  =
\begin{cases}
-(\mk - k) \;  G(k^\star), \qquad \text{for} \quad  k^\star < k \ll \mk - \frac{1}{L}   \\
-(\mk - k) \;  G(k), \qquad \text{for} \quad \frac{1}{L} \ll k < k^\star
\end{cases}
\end{eqnarray}

Thus for a deviation $k > k^\star$, the active droplet (in the inactive phase) has a microscopic width located around $L k^\star$ and the large deviation is still linear with minor corrections compared to the limiting case. For $k< k^\star$ then 
the system remains all the time in the inactive phase ($x_2 = 1$) and 
the droplet width shrinks leading to a non-linear large deviation cost in $k$.
The fact that larger droplet widths cannot be observed in the large deviations is due to the first order phase transition.

\medskip

\noindent
\emph{The very inactive regime:} 

For  $s\to\infty$ (\emph{i.e.} $\lambda\gg L$)
a single remaining site is active and one obtains the following
asymptotics~\cite{garrahan_dynamical_2007}:
\begin{align}
  \psi_L(s) &= -r_\infty+2ce^{-s}  + O(e^{-s})
  && \text{for}\; s\to\infty \text{ (\emph{i.e.} $\lambda \gg L$)}
\end{align}
where $r_\infty = 2c $ is the mean escape rate in the
configuration with one active site.
Performing the inverse Legendre transform, this implies that
\begin{equation}
\pi_L(k) = 
-r_\infty + L k \big(1-\log \frac{Lk}{2c}\big)
\qquad \text{for}\; k\ll L^{-1}
\end{equation}

\section{Conclusion }
\label{sec:conclusion}

 We have shown that the dynamical phase
coexistence occurring in the Fredrickson-Andersen model for histories at
small positive $s$ is well described by two Brownian interfaces
enclosing an island (or ``space-time
bubble''~\cite{chandler_dynamics_2010}) of activity of width $L^{\frac
  13}$ in a system of size $L$.
The scaling of this physical picture is reflected in
the finite size scaling of the dynamical free energy $\psi_L(s)$ of
the model.
We expect the same picture to be valid for 
a wider class of one-dimensional kinetically constrained models
where the particle number is not conserved.
It would be interesting to investigate the relation between our result and the statistics of large inactive bubbles
in the non-modified ($s=0$) dynamics.
\medskip

In general, the finite size scaling exponents depend on the dimension
and on the nature of the
constraints~\cite{bodineau_toninelli_2011}. Thus it would be
interesting to extend our study to more general dynamics, in
particular, to understand the interface fluctuations in higher
dimensional models.  This would be key to connect our approach to
realistic glass formers.
The quantitative link between those realistic (in general, atomistic)
models and kinetically constrained models have been explored in a
variety of studies~\cite{garrahan_geometrical_2002,berthier_spontaneous_2007,candelier_spatiotemporal_2010,keys_excitations_2011}.
Although it is beyond the scope of this article to use this
correspondence to provide quantitative predictions on activity large
deviation function in realistic models, we expect that the finite-size 
scaling exponents are related to geometrical features of inactive
regions (see~\cite{pitard_dynamic_2011} for an example of such exponent).

\begin{acknowledgements}
  We would like to thank Fr\'ed\'eric van Wijland for useful
  discussions, and Christophe Berthod and Thierry Giamarchi for the
  Mafalda cluster at DPMC, University of Geneva, where part of the
  simulations were run.  T.B., V.L. and C.T acknowledge funding from ANR SHEPI and C.T.
  from ERC Advanced Grant PTRELSS 228032.
\end{acknowledgements}

\appendix
\section{Scaling of $\psi_L(s)$ in the interface model}
\label{app:scaling_interface}

We determine in this appendix the finite size corrections to the
function $\hat\phi_L(\lambda)$, defined in~\eqref{eq:defphihat}, in
the effective interface description discussed in
part~\ref{subsec:interface}.
Let us first focus on a system with one boundary $x(t)$ between the
active and inactive regions.  It performs a random walk of jump rate
$p$ (resp. $q$) to the left (resp. right), starting from $x=0$ at time
$0$, and constrained to come back to $0$ at final time~$t$.
The walk $x(t)$ is reflected at $0$ so that $x(t)\geq 0$.
Appropriate values of $p$ and $q$ are discussed in
section~\ref{subsec: effective}. The computation is done for generic
values of $p$ and $q$.

 In the original model, the cost of maintaining the
interface (that is, for $x(\tau)$ to come back in $0$ at time
$\tau=t$) is given by the surface tension $\Sigma$.
In this effective description
\begin{equation}
  Z_{\text{eff}}(s,t)\equiv 
  \frac{\left\langle e^{-s\mk\int_0^t d\tau\:x(\tau)}\ \delta(x(t)=0)\right\rangle_{p,q}}{\big\langle \delta(x(t)=0)\big\rangle_{p,q}}
\end{equation}
where $\mk = 4 c^2(1-c)$ is the mean density of activity and
$\left\langle \cdot \right\rangle_{p,q}$ denotes the average on trajectories $x(\tau)_{0\leq\tau\leq t}$ without
constraint at final time. The normalisation $\left\langle
  \delta(x(t)=0)\right\rangle_{p,q}$ is e.g. fixed from $Z_{\text{eff}}(0,t)=1$.

\medskip

Let us denote by $P(x,X,t)$ the probability of being in $x$ at time $t$, having observed
a value $X$ of the area $\int_0^t d\tau\:x(\tau)$. The initial condition is $P(x,X,0)=\delta_{x,0}\delta(X)$.
Defining 
$\hat P(x,s,t)=\int dX\:e^{-s\mk X} P(x,X,t)$, one has (using the constraint at final time $x(t)=0$):
\begin{equation}
  Z_{\text{eff}}(s,t) = \frac{\hat P(0,s,t)}{\hat P(0,0,t)}
\end{equation}
Moreover, from Feynman-Kac formula, the time evolution of $\hat P(x,s,t)$ is given by
\begin{equation}
  \partial_t\hat P(x,s,t) = p \hat P(x+1,s,t)+q\hat P(x-1,s,t)-(p+q)\hat P(x,s,t) \ - \ s\mk x \hat P(x,s,t)
\end{equation}
and a reflection at $x=0$.
To symmetrize the walk, we now set $\hat Q(x,s,t)=(q/p)^{\frac x2}\hat P(x,s,t)$. One has again 
$
  Z_{\text{eff}}(s,t) = \frac{\hat Q(0,s,t)}{\hat Q(0,0,t)}
$
and the evolution of $\hat Q(x,s,t)$ writes
\begin{align}
  \partial_t\hat Q(x,s,t) = \sqrt{pq}\Big[ \hat Q (x+&1,s,t)+\hat Q(x-1,s,t)-2 \hat Q(x,s,t)\Big] 
  \nonumber \\ & 
  - s\mk x \hat Q(x,s,t) + \big(2\sqrt{pq}-p-q\big) \hat Q(x,s,t)
\end{align}
The normalisation has changed but we see that apart from the constant loss rate $\big(2\sqrt{pq}-p-q\big)$,
$\hat Q$ describes a \emph{symmetric} walk with the same term $ s\mk x \hat Q(x,s,t)$
corresponding to  weighting trajectories by the area $\int_0^t d\tau\:x(\tau)$. In other words, defining at last 
$\tilde Q(x,s,t) = \hat Q(x,s,t) e^{-t(2\sqrt{pq}-p-q)}$, one has
\begin{equation}
  Z_{\text{eff}}(s,t) = \frac{\tilde Q(0,s,t)}{\tilde Q(0,0,t)}
\end{equation}
and from the equation of evolution
\begin{equation}
  \partial_t\tilde Q(x,s,t) = \sqrt{pq}\Big[ \tilde Q (x+1,s,t)+\tilde Q(x-1,s,t)-2 \tilde Q(x,s,t)\Big] 
  - s\mk x \tilde Q(x,s,t) 
\end{equation}
we see that $\tilde Q(0,0,t)$ does not increase exponentially in time since for $s=0$ the equation
describes a simple symmetric random walk. Moreover, in the large size limit ($s\to 0$), the evolution of $\tilde Q(x,s,t)$
is governed by the continuous in space operator
\begin{equation}
   \sqrt{pq}\partial_x^2 - s\mk x 
  = 2 \sqrt{pq}\: \Big[ \frac 12 \partial_x^2 - \frac{s\mk}{2\sqrt{pq}} x  \Big]
\end{equation}
with reflecting boundary condition in $0$.
For bridges, the spectrum is known and its largest eigenvalue is given by~\cite{majumdar_airy_2005}
\begin{equation}
  \psi^\text{non-per}_{\text{eff}}(s)=2\sqrt{pq} \left(\frac{s\mk}{2\sqrt{pq}}\right)^{\frac 23} 2^{-\frac 13}\alpha_1
\end{equation}
where $\alpha_1\approx 2.3381...$ is the first zero of the Airy
function on the negative real axis.  In periodic boundary conditions,
one has two interfaces and the equivalent jump rates are multiplied by
$2$. Finally, this gives in the $s\to 0$ limit:
\begin{equation}
  \psi^\text{per}_{\text{eff}}(s)=4\sqrt{pq} \left(\frac{s\mk}{4\sqrt{pq}}\right)^{\frac 23} 2^{-\frac 13}\alpha_1
\end{equation}

\section{A generic identity between large deviation functions}
\label{app:identity_ldfKR}

In this appendix we prove an identity used in
part~\ref{sec:numerical_results_cloning_density} between large deviation
functions associated to the activity $K_t$ and to the escape rate $R_t$
defined below.
We consider a Markov process on a finite number of configurations
$\{\C\}$, with transition rates $W(\C\to\C')$ between
configurations. The activity $K_t$ is an history-dependent observable
increasing by $1$ upon jumping from $\C$ to $\C'$.
The probability $P(\C,K,t)$ of being in $\C$ at time $t$ having
observed a value $K$ of the observable $K_t$ thus evolves in time through
\begin{equation}
  \partial_t P(\C,K,t) = 
  \sum_{\C'} W(\C'\to\C) P(\C',K-1,t) - r(\C) P(\C,K,t)
\end{equation}
with $r(\C)=  \sum_{\C'} W(\C\to\C')$ the escape rate from configuration $\C$.
The Laplace transform $P(\C,s,t) =\sum_K e^{-sK}P(\C,K,t)$ verifies
\begin{equation}
  \partial_t P(\C,s,t) = 
  \sum_{\C'} e^{-s}W(\C'\to\C) P(\C',s,t) - r(\C) P(\C,s,t)
\end{equation}
The cumulant generating function $\psi(s)$  defined in the infinite time
limit as $\langle e^{-sK_t}\rangle\sim e^{t\psi(s)}$ is the largest eigenvalue of the operator
$\mathbb W_s$ of elements
\begin{equation}
  \big(\mathbb W_s\big)_{\C\C'}= e^{-s} W(\C'\to\C) - r(\C)\delta_{\C\C'}
 \label{eq:defWofs_generic}
\end{equation}
since $\langle e^{-sK_t}\rangle=\sum_\C P(\C,s,t) $.
Both $K_t$ and $R_t=\int_0^td\tau\:r(\C(\tau))$ quantify the activity
of the histories. Their large deviation functions (ldf) are closely
related.  Indeed, let's consider the joint ldf
\begin{equation}
  \Psi(s,\sigma)= \lim_{t\to\infty}\frac 1t \log \big\langle e^{-sK_t-\sigma R_t}\big\rangle
\end{equation}
As previously, one checks that $\Psi(s,\sigma)$ is given by the
maximum eigenvalue of the operator $\mathbb W_{s,\sigma}$ of
elements~\cite{garrahan_first-order_2009}
\begin{equation}
 \big(\mathbb W_{s,\sigma}\big)_{\C\C'}= e^{-s} W(\C'\to\C) - (1+\sigma)r(\C)\delta_{\C\C'}
\end{equation}
It verifies the symmetry $\mathbb W_{s,\sigma} = (1+\sigma)\mathbb W_{s+\log(1+\sigma),0}$ and so does the ldf:
\begin{equation}
 \Psi(s,\sigma)= (1+\sigma)\Psi(s+\log(1+\sigma),0)
 \label{eq:symmetryPsissigma}
\end{equation}
Besides, the mean values of $K_t$ and $R_t$ in the $s$-state are given by
\begin{equation}
  \frac 1t \langle K_t\rangle_s = -\partial_s\Psi(s,\sigma)\big|_{\sigma=0}
\qquad
  \frac 1t \langle R_t\rangle_s = -\partial_\sigma\Psi(s,\sigma)\big|_{\sigma=0}
\end{equation}
Differentiating the symmetry~\eqref{eq:symmetryPsissigma} with respect to $\sigma$ and sending $\sigma$ to 0, one gets
\begin{equation}
   \partial_\sigma\Psi(s,\sigma)\big|_{\sigma=0}
   =
   \Psi(s,0)+ \partial_s\Psi(s,\sigma)\big|_{\sigma=0}
\end{equation}
which implies
\begin{equation}
 { \psi(s) = \frac 1t \langle K_t\rangle_s - \frac 1t \langle R_t\rangle_s }
 \label{eq:symmetrypsiKR}
\end{equation}
 This relation is generic. It leads to the relation~\eqref{eq:KtRt} between
scaling exponents for the FA model  in the inactive regime
$s=\lambda/L$.

\section{Bernoulli approximation to determine $\phi_L(\lambda)$}
\label{app:Bernoulli}

In this appendix, we obtain the expression of the Courant-Fischer
optimisation principle of part~\ref{sec:numerical_results_Bernoulli}
for Bernoulli states.
Before this, one needs to symmetrize the evolution operator $\mathbb
W_s$ introduced in~\eqref{eq:defWofs_generic}.  We take the notation
of appendix~\ref{app:identity_ldfKR}.
Assuming that the jump rates verify the detailed balance symmetry
$W(\C\to\C')P_{\text{eq}}(\C)=W(\C'\to\C)P_{\text{eq}}(\C')$, it is
generically possible to symmetrize the operator of evolution $\mathbb
W_s$ through the similarity transformation $\mathbb W^\sym_s \equiv
\hat P^{-\frac 12}_{\text{eq}} \mathbb W_s \hat P^{\frac
  12}_{\text{eq}} $, where $\hat P_{\text{eq}}$ is the diagonal
operator of elements $P_{\text{eq}}(\C)$.
 Upon symmetrisation,
we have that $\psi(s)$ is also the largest eigenvalue of the symmetric operator
$\mathbb W^\sym_s$ of elements
\begin{equation}
  \big(\mathbb W^\sym_s\big)_{\C\C'}= e^{-s} \big[W(\C'\to\C) W(\C\to\C') \big]^{\frac 12} - r(\C)\delta_{\C\C'}
\end{equation}

It is convenient to represent the operator of evolution in terms of
spin $\frac 12$ operators $\sigma^\pm$ and $\hat n$.  On each site
$i$, $\sigma^\pm$ is the creation/annihilation operator and $\hat n$
is the counting operator.  They are defined by
\begin{align}
  \sigma^+|0\rangle &= |1\rangle &    \sigma^-|0\rangle &= 0         & \hat n |0\rangle &= 0
\\
  \sigma^+|1\rangle &= 0         &    \sigma^-|1\rangle &= |0\rangle & \hat n |1\rangle &= |1\rangle
\end{align}
Where $|0\rangle$ and $|1\rangle$ are the vectors for empty and occupied states.
One has
\begin{equation}
  \mathbb W_s = \sum_{1\leq i\leq L} \Big[e^{-s}\big(c\sigma^+_i+(1-c)\sigma_i^-\big)-c(1-\hat n_i)-(1-c)\hat n_i\Big](\hat n_{i+1}+\hat n_{i-1})
\end{equation}
Transition rates obey detailed balance with respect to the product Bernoulli measure
of uniform density $c$ (conditioned to exclude the fully inactive
configuration). The symmetrized operator of evolution writes
\begin{equation}
  \mathbb W^\sym_s = \sum_{1\leq i\leq L} \big[e^{-s}\sqrt{c(1-c)}(\sigma^+_i+\sigma_i^-)-c(1-\hat n_i)-(1-c)\hat n_i\big](\hat n_{i+1}+\hat n_{i-1})
\end{equation}

Defining the vector $|\rho\rangle$ corresponding to the Bernoulli distribution of density~$\sqrt{\rho}$:
\begin{equation}
  |\rho\rangle = \sqrt{\rho} |1\rangle + (1-\sqrt{\rho}) |1\rangle
\end{equation}
and taking $|X\rangle$ in~\eqref{eq:phiL_optim} to be the state
$|\rho_1\ldots\rho_L\rangle^{\text{cond}}$ conditioned to exclude the
fully empty configuration:
\begin{equation}
    |\rho_1\ldots\rho_L\rangle^{\text{cond}} 
=   \sum_{n_i:\sum_in_i\neq0}\langle n_1\ldots n_L|\rho_1 \ldots \rho_L\rangle \:|n_1\ldots n_L\rangle
\end{equation}
one obtains~\eqref{eq:psiBern} by direct computation.

\bibliography{psiK_FA_finite-size}{}
\bibliographystyle{plain_url}

\end{document}